# COSMOLOGY WITH BOUNCE BY FLAT SPACE-TIME THEORY OF GRAVITATION AND A NEW INTERPRETATION

Walter Petry
Mathematisches Institut der Universitaet Duesseldorf, D-40225 Duesseldorf
E-mail: wpetry@meduse.de
petryw@uni-duesseldorf.de

**Abstract.** General relativity predicts a singularity in the beginning of the universe being called big bang. Recent developments in loop quantum cosmology avoid the singularity and the big bang is replaced by a big bounce. A classical theory of gravitation in flat space-time also avoids the singularity under natural conditions on the density parameters. The universe contracts to a positive minimum and then it expands during all times. It is not symmetric with regard to its minimum implying a finite age measured with proper time of the universe. The space of the universe is flat and the total energy is conserved. Under the assumption that the sum of the density parameters is a little bit bigger than one the universe is very hot in early times. Later on, the cosmological model agrees with the one of general relativity. A new interpretation of a non-expanding universe may be given by virtue of flat space-time theory of gravitation.

## 1. Introduction

Einstein's theory of general relativity is generally accepted as the most powerful theory of gravitation by virtue of its well-known predictions. It gives a singularity in the beginning of the universe being called big bang and which has been accepted for long times. But recent developments of loop quantum cosmology avoid the singularity and it is replaced by a big bounce. There are many authors who have studied the big bounce by the use of loop quantum cosmology, see e.g. [1-5] and the extensive references therein. One compare also the popular book [6] on this subject. A big bounce in the beginning of the universe has already been studied by Priester (see e.g. [7] ). Observational hints on a big bounce can be found in [8].

In 1981 the author [9] has studied a covariant theory of gravitation in flat space-time. There exists an extensive study of this theory since that time. The energy-momentum of gravitation is a covariant tensor and the total energy-momentum of all kinds of matter and fields including that of gravitation is the source of the gravitational field. The total energy-momentum is conserved. The theory gives the same results as general relativity to the accuracy demanded by the experiments for: gravitational redshift, light deflection, perihelion precession, radar time delay, post-Newtonian approximation, gravitational radiation, and the precession of the spin axis of a gyroscope in the orbit of a rotating body. But there exist also differences to the results of general relativity, these are: the theorem of Birkhoff doesn't hold and the theory gives non-singular cosmological models (no big bang). A summary of flat space-time theory of gravitation with the mentioned results can be found in [10] where also references to the detailed studies are given. Non-singular cosmological models studied by the use of flat space-time theory of gravitation can be found e.g. in the papers [11-15].

Subsequently, we follow along the lines of the above mentioned articles. Let us assume a hmogeneous, isotropic universe consisting of matter, radiation and dark energy, given by a cosmological constant. The theory of gravitation in flat space-time implies a flat space. Under the assumption that the sum of all the density parameters is bigger than one, the solution describing the universe is non-singular, i.e. all the energies are finite. The universe contracts to a minimum and then it expands for all times. The sum of all the energies of matter, radiation, dark energy, and the gravitational energy is conserved. Assuming that the sum of all the density parameters is a little bit bigger than one then the universe becomes very hot in early times. The time where the contracting universe enters into the expanding one corresponds to the big bang of Einstein's theory. Some time after this point of contraction to expansion the solution agrees with the result of Einstein's general theory of relativity. There is no need of inflation because the space of the universe is always flat. It is worth to mention that the theory of gravitation indicates that an other interpretation as contracting and expanding universe is possible. The universe is non-stationary and the time dependence follows by the transformation of the different kinds of energy into one another whereas the total energy is conserved. This transformation of the energies is also the reason for the observed redshifts at distant galaxies. This interpretation also solves the problem of velocities higher than the light velocity at very distant galaxies.

## 2. Gravitation In Flat Space-Time

In this section the subsequently used covariant theory of gravitation in flat space-time [9] is shortly summarized. The line-element of flat space-time is

$$(ds)^2 = -\eta_{ij} dx^i dx^j \qquad (2.1)$$

where $(\eta_{ij})$ is a symmetric tensor. In the special case where $(x^1, x^2, x^3)$ are the Cartesian coordinates, $x^4 = ct$ and

$$(\eta_{ij}) = diag(1,1,1,-1) \qquad (2.2)$$

the space-time metric (2.1) is the pseudo-Euclidean geometry. Put

$$\eta = \det(\eta_{ij}). \qquad (2.3)$$

The gravitational field is desribed by a symmetrric tensor $(g_{ij})$. Let $(g^{ij})$ be defined by

$$g_{ik} g^{kj} = \delta_i^{\ j}, \quad g^{ik} g_{kj} = \delta^i_{\ j} \qquad (2.4)$$

and put analogously to (2.3)

$$G = \det(g_{ij}). \qquad (2.5)$$

The proper time $\tau$ is defined similarly to (2.1) by the quadratic form

$$c^2 (d\tau)^2 = -g_{ij} dx^i dx^j. \qquad (2.6)$$

The Lagrangian of the gravitational field is given by

$$L_G = -\left(\frac{-G}{-\eta}\right)^{1/2} g_{ij} g_{kl} g^{mn} \left( g^{ik}_{\ /m} g^{jl}_{\ /n} - \frac{1}{2} g^{ij}_{\ /m} g^{kl}_{\ /n} \right) \qquad (2.7)$$

where the bar / denotes the covariant derivative relative to the flat space-time metric (2.1). The Lagrangian of the dark energy ( given by the cosmological constant $\Lambda$ ) has the form

$$L_\Lambda = -8\Lambda \left(\frac{-G}{-\eta}\right)^{1/2}. \qquad (2.8)$$

Put

$$\kappa = 4\pi k / c^4 \qquad (2.9)$$

where $k$ denotes the gravitational constant, then the mixed energy-momentum tensors of the gravitational field, of dark energy and of matter of a perfect fluid are given by the following expressions

$$\overset{G}{T}{}^i_{\ j} = \frac{1}{8\kappa} \left[ \left(\frac{-G}{-\eta}\right)^{1/2} g_{kl} g_{mn} g^{ir} \left( g^{km}_{\ /j} g^{\ln}_{\ /r} - \frac{1}{2} g^{kl}_{\ /j} g^{mn}_{\ /r} \right) + \frac{1}{2} \delta^i_{\ j} L_G \right] \qquad (2.10a)$$

$$\overset{\Lambda}{T}{}^i_{\ j} = \frac{1}{16\kappa} \delta^i_{\ j} L_\Lambda \qquad (2.10b)$$

$$\overset{M}{T}{}^i_{\ j} = (\rho + p) g_{jk} u^i u^k + \delta^i_{\ j} p c^2. \qquad (2.10c)$$

Here, $\rho, p$ and $(u^i)$ denote density, pressure and four-velocity $\left(\dfrac{dx^i}{d\tau}\right)$ of matter. It holds by virtue of (2.6)

$$c^2 = -g_{ij} u^i u^j. \qquad (2.11)$$

Define the covariant differential operator

$$\tilde{R}^i_{\ j} = \left[ \left(\frac{-G}{-\eta}\right)^{1/2} g^{kl} g_{jm} g^{mi}_{\ /l} \right]_{/k} \qquad (2.12)$$

of order two in divergence form, then the field equations for the potentials $(g_{ij})$ can be written in the covariant form

$$\tilde{R}^i_{\ j} - \frac{1}{2} \delta^i_{\ j} \tilde{R}^k_{\ k} = 4\kappa T^i_{\ j} \qquad (2.13)$$

where

$$T^i_{\ j} = \overset{G}{T}{}^i_{\ j} + \overset{\Lambda}{T}{}^i_{\ j} + \overset{M}{T}{}^i_{\ j} \qquad (2.14)$$

is the total energy-momentum tensor. The equations of motion are given in covariant form by

$$\overset{M}{T}{}^{k}{}_{i/k} = \frac{1}{2} g_{kl/i} \overset{M}{T}{}^{kl} \tag{2.15}$$

where

$$\overset{M}{T}{}^{ij} = g^{jk} \overset{M}{T}{}^{i}{}_{k} \tag{2.16}$$

is the symmetric energy-momentum tensor of matter. In addition to the field equations (2.13) and the equations of motion (2.15) the conservation law

$$T^{k}{}_{i/k} = 0 \tag{2.17}$$

of the total energy-momentum tensor holds. All the equations (2.13), (2.15) and (2.17) are generally covariant and the energy-momentum (2.10a) of the gravitational field is a tensor in contrast to that of general relativity.

The field equations (2.13) are formally similar to the equations of Einstein's general relativity theory but $\tilde{R}^{i}{}_{j}$ is not the Ricci tensor and the source of the gravitational field includes the energy-momentum tensor of gravitation. It is worth to mention that the field equations (2.13) together with the equations (2.15), respectively (2.17) imply the equations (2.17), respectively (2.15).

### 3. Isotropic Cosmological Model

In this section the flat space-time theory of gravitation is applied to homogeneous, isotropic cosmological models. The pseudo-Euclidean geometry (2.1) with (2.2) is assumed. The matter tensor is given by (2.10c) with

$$u^{i} = 0 \quad (i = 1,2,3) \tag{3.1}$$

and

$$p = p_m + p_r, \quad \rho = \rho_m + \rho_r \tag{3.2}$$

where the indices $m$ and $r$ denote matter and radiation respectively. The equations of state for matter (dust) and radiation are

$$p_m = 0, \quad p_r = \rho_r / 3. \tag{3.3}$$

By virtue of (3.1) the potentials are give by

$$\begin{aligned} g_{ij} &= a(t)^2, \quad (i = j = 1,2,3) \\ &= -1/h(t), \quad (i = j = 4) \\ &= 0, \quad (i \neq j) \end{aligned} \tag{3.4}$$

where all the functions depend only on $t$ by virtue of the homogeneity of the model. The four-velocity (2.11) has by the use of (3.1) and (3.4) the form

$$(u^i) = (0,0,0, ch^{1/2}). \tag{3.5}$$

For the present time $t_0 = 0$ the following initial conditions are assumed:

$$a(0) = h(0) = 1, \quad \dot{a}(0) = H_0, \quad \dot{h}(0) = \dot{h}_0, \quad \rho_m(0) = \rho_{m0}, \quad \rho_r(0) = \rho_{r0} \tag{3.6}$$

where the dot denotes the time-derivative, $H_0$ is the well-known Hubble constant and $\dot{h}_0$ is a constant which doesn't appear by Einstein's theory. The constants $\rho_{m0}$ and $\rho_{r0}$ denote the present densities of matter and radiation. It is worth to mention that general relativity implies $h(t) \equiv 1$ which is not possible in flat space-time theory of gravitation. This will be important to avoid the singularity.

Under the assumption that matter and radiation do not interact the equations of motion (2.15) can be solved by the use of (3.1) to (3.5). It follows

$$\rho_m = \rho_{m0} / h^{1/2}, \quad p_m = 0, \quad \rho_r = 3 p_r = \rho_{r0} / (ah^{1/2}). \tag{3.7}$$

The field equations (2.13) with (2.10) to (2.14) imply by (3.1) to (3.5) the two non-linear differential equations:

$$\frac{d}{dt}\left(a^3 h^{1/2} \frac{\dot{a}}{a}\right) = 2\kappa c^4 \left[\frac{1}{2}\rho_m + \frac{1}{3}\rho_r + \frac{\Lambda}{2\kappa c^2} \frac{a^3}{h^{1/2}}\right] \tag{3.8a}$$

$$\frac{d}{dt}\left(a^3 h^{1/2} \frac{\dot{h}}{h}\right) = 4\kappa c^4 \left[\frac{1}{2}\rho_m + \rho_r + \frac{1}{8\kappa c^2} L_G - \frac{\Lambda}{2\kappa c^2} \frac{a^3}{h^{1/2}}\right]. \tag{3.8b}$$

Here, it holds

$$L_G = \frac{1}{c^2} a^3 h^{1/2} \left( -6\left(\frac{\dot{a}}{a}\right)^2 + 6\frac{\dot{a}}{a}\frac{\dot{h}}{h} + \frac{1}{2}\left(\frac{\dot{h}}{h}\right)^2 \right) \tag{3.9}$$

where the gravitational energy is given by $\frac{1}{16\kappa} L_G$.

The proper time is:

$$c^2 (d\tau)^2 = -a^2 \left( (dx^1)^2 + (dx^2)^2 + (dx^3)^2 \right) + \frac{1}{h}(ct)^2. \tag{3.10}$$

The conservation of the total energy has the form:

$$(\rho_m + \rho_r)c^2 + \frac{1}{16\kappa} L_G + \frac{\Lambda}{2\kappa}\frac{a^3}{h^{1/2}} = \lambda c^2 \tag{3.11}$$

where $\lambda$ is a constant of integration. The equations (3.8) and (3.9) give by the use of (3.11) and the initial conditions (3.6):

$$\frac{\dot{h}}{h} = -6\frac{\dot{a}}{a} + 2\frac{4\kappa c^4 \lambda t + \varphi_0}{2\kappa c^4 \lambda t^2 + \varphi_0 t + 1} \tag{3.12}$$

with

$$\varphi_0 = 3H_0 \left( 1 + \frac{1}{6}\frac{\dot{h}_0}{h_0} \right). \tag{3.13}$$

Integration of relation (3.12) yields:

$$a^3 h^{1/2} = 2\kappa c^4 \lambda t^2 + \varphi_0 t + 1. \tag{3.14}$$

Equation (3.11) gives at present time $t_0 = 0$ by the use of (3.6) and (3.7)

$$\frac{1}{3}\left(8\kappa c^4 \lambda - \varphi_0^2\right) = 4\left[ \frac{8}{3}\pi k \left( \rho_{m0} + \rho_{r0} + \frac{\Lambda c^2}{8\pi k} \right) - H_0^2 \right]. \tag{3.15}$$

It follows from (3.11) by the use of (3.9), (3.12), (3.14), (3.7) and (3.15) the formula

$$\left(\frac{\dot{a}}{a}\right)^2 = \frac{1}{\left(2\kappa c^4 \lambda t^2 + \varphi_0 t + 1\right)^2} \left( -\left(\frac{8}{3}\pi k \left( \rho_{m0} + \rho_{r0} + \frac{\Lambda c^2}{8\pi k} \right) - H_0^2 \right) + \frac{\Lambda c^2}{3}a^6 + \frac{8}{3}\pi k \rho_{m0} a^3 + \frac{8}{3}\pi k \rho_{r0} a^2 \right). \tag{3.16}$$

Let us introduce the density parameters:

$$\Omega_m = \frac{8\pi k \rho_{m0}}{3H_0^2}, \quad \Omega_r = \frac{8\pi k \rho_{r0}}{3H_0^2}, \quad \Omega_\Lambda = \frac{\Lambda c^2}{3H_0^2} \tag{3.17}$$

and define

$$K_0 = (\Omega_m + \Omega_r + \Omega_\Lambda - 1)/\Omega_m \tag{3.18}$$

then the differential equation (3.16) can be rewritten in the form:

$$\left(\frac{\dot{a}}{a}\right)^2 = \frac{H_0^2}{\left(2\kappa c^4 \lambda t^2 + \varphi_0 t + 1\right)^2} \left[ -\Omega_m K_0 + \Omega_r a^2 + \Omega_m a^3 + \Omega_\Lambda a^6 \right] \tag{3.19a}$$

with the initial condition

$$a(0) = 1. \tag{3.19b}$$

Hence, a solution of (3.19) together with (3.14) describes a homogeneous, isotropic cosmological model in flat space-time theory of gravitation.

### 4. Cosmology With A Bounce

In this section we will study the solutions of (3.19) with (3.14) and show that non-singular cosmological models with a bounce exist.

Relation (3.15) can be rewritten by the use of the density parameters (3.17) and the definition (3.18)

$$\frac{8\kappa c^4 \lambda}{H_0^2} - \left(\frac{\varphi_0}{H_0}\right)^2 = 12\Omega_m K_0. \tag{4.1}$$

A necessary condition to avoid a singular solution of (3.19) is the condition

$$K_0 > 0. \tag{4.2}$$

Inequality (4.2) is by the use of (4.1) equivalent to

$$2\kappa c^4 \lambda t^2 + \varphi_0 t + 1 > 0 \tag{4.3}$$

for all $t \in R$. Then, the differential equation (3.19) has a positive solution $a(t)$ and relation (3.14) gives a positive function $h^{1/2}(t)$. Therefore, condition (4.2) is necessary and sufficient for non-singular cosmological models by virtue of (3.7). Then, there exists a time $t_1$ such that

$$\dot{a}(t_1) = 0. \tag{4.4}$$

Put $a_1 = a(t_1)$, then the differential equation (3.19a) implies at $t = t_1$:

$$\Omega_r a_1^2 + \Omega_m a_1^3 + \Omega_\Lambda a_1^6 = \Omega_m K_0 \tag{4.5}$$

and for all $t \in R$

$$a(t) \geq a_1 > 0. \tag{4.6}$$

The assumption

$$a_1 \ll a(0) = 1 \tag{4.7}$$

gives by virtue of (4.5)

$$K_0 \ll 1. \tag{4.8}$$

Hence, it follows by (3.18)

$$\Omega_r + \Omega_m + \Omega_\Lambda = 1 + \Omega_m K_0, \tag{4.9}$$

i.e. the sum of all the density parameters is a little bit bigger than one. The condition (4.7) is also iumportant for a very hot universe in the beginning because the temperature is given by

$$T(t) = T_0 / a(t) \tag{4.10}$$

where $T_0$ is the present temperature of radiation. For $t < t_1$ the universe contracts and then it expands as $t > t_1$. Hence, there exists a bounce in the early universe.

The differential equation (3.19) is written in the case $t_1 \leq t$ (expanding universe) in the form:

$$\frac{\dot{a}}{a} = \frac{H_0}{2\kappa c^4 \lambda t^2 + \varphi_0 t + 1} \left(-\Omega_m K_0 + \Omega_r a^2 + \Omega_m a^3 + \Omega_\Lambda a^6\right)^{1/2}$$
$$a(0) = 1 \tag{4.11}$$

and for $t \leq t_1$ (contracting universe) in the form:

$$\frac{\dot{a}}{a} = -\frac{H_0}{2\kappa c^4 \lambda t^2 + \varphi_0 t + 1} \left(-\Omega_m K_0 + \Omega_r a^2 + \Omega_m a^3 + \Omega_\Lambda a^6\right)^{1/2}$$
$$a(t_1) = a_1. \tag{4.12}$$

In the special case $\Omega_r = 0$ an analytic solution of the expanding universe (4.11) can be given (see e.g. [12]):

$$a^3(t) = 2K_0 / \left(1 - (1 - 2K_0)\cos(\sqrt{3}\alpha(t)) - 2(K_0/\Omega_m)^{1/2} \sin(\sqrt{3}\alpha(t))\right) \tag{4.13a}$$

where

$$\alpha(t) = arctg\left((3\Omega_m K_0)^{1/2} H_0 t / \left(1 + \frac{1}{2}\varphi_0 t\right)\right). \tag{4.13b}$$

For the subsequent considerations compare [14].

The time $t_1$ is given by (in the case of $\Omega_r = 0$)

$$H_0 t_1 = -1 / \left(\frac{1}{2}\frac{\varphi_0}{H_0} - \frac{(3\Omega_m K_0)^{1/2}}{A}\right) \tag{4.14a}$$

with

$$A = tg\left(\frac{1}{\sqrt{3}}\left(-\pi + arctg\left(2\frac{(K_0/\Omega_m)^{1/2}}{1-K_0}\right)\right)\right) \approx 4.0338 + O(K_0^{1/2}).\tag{4.14b}$$

Relation (4.13) gives two different kinds of solutions (see [14]):
(1) The denominator of (4.13) is positive and vanishes as $t \to \infty$ implying

$$\frac{1}{2}\frac{\varphi_0}{H_0} \approx \frac{3}{2}\Omega_m/\left(1-\sqrt{\Omega_\Lambda}\right)\tag{4.15}$$

where expressions containing $K_0$ are omitted by virtue of (4.8).
(2) The denominator of (4.13) is always positive as $t \to \infty$ implying

$$\frac{1}{2}\frac{\varphi_0}{H_0} > \frac{3}{2}\Omega_m/\left(1-\sqrt{\Omega_\Lambda}\right).\tag{4.16}$$

In both cases the function $a(t)$ is increasing after the time $t_1$. In the first case $a(t)$ converges to infinity as $t$ goes to infinity whereas in the second case the function $a(t)$ converges to a finite value as $t$ goes to infinity. Subsequently, we will only consider the interesting case (1). Condition (4.15) is by the use of (3.13) a condition on the initial value $\dot{h}_0$ of (3.6).

Let us define the time $\tilde{t}_1$ by

$$H_0\tilde{t}_1 = -\frac{1}{2}\frac{\varphi_0}{H_0}\Big/\left(\left(\frac{1}{2}\frac{\varphi_0}{H_0}\right)^2 + 3\Omega_m K_0\right)\tag{4.17}$$

then, for $\tilde{t}_1 \ll t$ the solution (4.13a) can be approximated by

$$a^3(t) \approx \left(\frac{3}{2}\frac{\Omega_m}{1-\sqrt{\Omega_\Lambda}}\right)^2 (H_0 t - H_0\tilde{t}_1)^2 / \left[3\sqrt{\Omega_\Lambda}(H_0 t - H_0\tilde{t}_1) + \left(1-\sqrt{\Omega_M}\right)^2/\Omega_m\right]\tag{4.18a}$$

and by the use of (3.14)

$$h^{1/2}(t) \approx 3\sqrt{\Omega_\Lambda}(H_0 t - H_0\tilde{t}_1) + \left(1-\sqrt{\Omega_\Lambda}\right)^2/\Omega_m.\tag{418b}$$

For $t \leq t_1$ the function $a(t)$ of the differential equation (4.12) starts at

$$a(-\infty) \approx \left(2/(1-\cos(\sqrt{3}\pi))\right)^{1/3} a_1 \approx 1.8161 a_1\tag{4.19}$$

and decreases as $t \to t_1$ to $a_1$.

The proper time $\tau$ from the beginning of the universe is given by

$$\tau(t) = \int_{-\infty}^{t} 1/h^{1/2}(t)dt.\tag{4.20}$$

The proper time $\tau(t_1)$, i.e. from the beginning of the universe till $t_1$, is finite by virtue of (3.14), (4.19) and (4.7). The proper time $\tau(t)$ of the universe is by (4.20) increasing with increasing $t$ and goes to infinity as $t$ goes to infinity by virtue of (4.18b). It follows for $t \gg \tilde{t}_1$ the proper time

$$\tau(t) = \tau_0 + \frac{1}{3\sqrt{\Omega_\Lambda}}\frac{1}{H_0}\ln\left(h^{1/2}(t)\right)\tag{4.21a}$$

with a suitable constant $\tau_0$. Therefore, it holds for $\tau(t)$ sufficiently large

$$h^{1/2}(t) = \exp\left(3\sqrt{\Omega_\Lambda}H_0(\tau(t)-\tau_0)\right).\tag{4.21b}$$

Hence, under the condition (4.15), the function $a(t)$ starts by virtue of (4.19) from a small positive value $a(-\infty) > a_1$ and decreases to $a_1$ as $t \to t_1$. Then, $a(t)$ increases for all times $t > t_1$ and goes to infinity as $t \to \infty$. It is worth to mention that $a(t)$ is not symmetric with regard to its minimum at $t_1$.

Let us now introduce the proper time (4.20) in the differential equation (3.19). It follows by the use of (3.14)

$$\frac{1}{a^2}\left(\frac{da}{d\tau}\right)^2 = H_0^2\left(-\frac{\Omega_m K_0}{a^6} + \frac{\Omega_r}{a^4} + \frac{\Omega_m}{a^3} + \Omega_\Lambda\right)$$
$$a(0)=1.$$
(4.22)

Hence, this differential equation is by the use of (4.8) and (4.2) for $a >> a_1$ identical with the differential equation of general relativity describing a homogeneous, isotropic universe. Therefore, all the results of general relativity are valid. But for sufficiently small $a(t)$, the solution is quite different from that of general relativity and it has no singularity in contrast to Einstein's theory.

### 5. A New Interpretation

In this section a new interpretetion of the results of section 4 are given. All the formulae and results of the previous chapter are valid exept the interpretation of a bounce, i.e. of a collapsing and expanding universe. This is possible by virtue of the conservation of the total energy (3.11) with (3.7) and (3.9) in flat space-time theory of gravitation. The universe can be interpreted as non-expanding where the redshift follows by the transformation of the different kinds of energy into one another (see [14], [15] ). The formula for the redshift is identical with the one of the expanding universe. The derivation of this result can be found in [13,14,15].

In the beginning of the non-expanding universe, no matter, no radiation and no dark energy exist by virtue of (3.14) and (4.19). In the course of time radiation, matter and dark energy arise whereas the total energy is conserved. Later on, matter and radiation decrease (this corresponds to the expanding universe). Formula (3.7) for matter with (4.21b) implies that matter is exponentially decaying for sufficiently large proper time $\tau$ in the non-expanding universe analogously to the radioactive decay whereas dark energy increases to a finite value as $t$, respctively $\tau$ goes to infinity. It is worth to mention that this result depends on the assumed dark energy given by a cosmological constant.

It seems that this result is a more natural interpretation of the universe implied by the use of gravitation in flat space-time.The problem of velocities of galaxies higher than light velocity doesn't arise and something like inflation is superfluous because in the beginning of the universe, no matter, no radiation and no dark energy exist , i.e. all the energy is in form of gravitation and the space is flat for all times.